\def \cm{~\rm{cm}}

\def \K{~\rm{K}}
\def \g{~\rm{g}}

\def \erg{~\rm{erg}}
\def \yrs{~\rm{yrs}}
\def \yr{~\rm{yr}}

\documentclass[12pt,preprint]{aastex}
\usepackage{epsfig}
\usepackage{graphics}
\usepackage{rotate}

\begin{document}

\title{Stellar structure and mass loss during the early 
post asymptotic giant branch}

\author{Noam Soker and Amos Harpaz}

\affil{Department of Physics, Oranim, Tivon 36006, Israel}

\begin{abstract}

Late asymptotic giant branch (AGB) and early post-AGB stars which
are progenitors of planetary nebulae lose mass at extremely
high rate, in what is termed a superwind. 
We show that the existence of this superwind during the post-AGB 
phase cannot be explained with models where the stellar effective 
temperature is the sole main physical parameter which determine
the mass loss rate.
Instead, we argue that the envelope structure, in particular
the entropy and density gradients, should be among the main
parameters which determine the mass loss rate on the tip of
the AGB and the early post-AGB evolutionary phases. 
The entropy profile becomes steeper and the density profile becomes
shallower as the star becomes hotter on the early post-AGB phase,
until the star heats-up to $T \gtrsim 8000\K$.
We propose that mass loss rate stays very high because of the
envelope structure, and drops only when the effects
of the temperature becomes important once again as the post-AGB
star heats to $\sim 6,000 \K$.
We do not propose a new mass loss mechanism, but rather mention 
several mechanisms by which these profiles may influence the mass 
loss rate within the popular mechanism for mass loss on the AGB, 
where pulsations coupled with radiation pressure on dust 
cause the high mass loss rate. 

\end{abstract}

\keywords{
stars: AGB and post-AGB -- circumstellar matter -- stars: mass loss --
planetary nebulae: general  }
\section {INTRODUCTION}

Models for the acceleration of winds from asymptotic giant branch (AGB)
stars which are based on radiation pressure on dust agree well with
observations (e.g., Elitzur,  Ivezic \& Vinkovic 2002). 
The exact mass loss rate depends on the process which initiates
the outflow (Elitzur et al. 2002).
One of the most popular mechanism to initiate the outflow, 
which is based both on observations and theoretical studies, 
is stellar pulsation
(e.g., Wood 1979; Bowen 1988 H\"ofner et al. 1996; H\"ofner 1999;
Winters et al.\ 2000; Andersen, H\"ofner \& Gautschy-Loidl 2002).
In such models, it is found that the mass loss rate substantially
increases as the star evolves along the AGB
(e.g., Bedijn 1988; Bowen \& Willson 1991; H\"ofner \& Dorfi 1997; 
Wachter et al.\ 2002, hereafter WA02),
with strong dependence on the stellar mass, effective temperature,
and luminosity (or radius).
This behavior along the AGB is in agreement with observations
which require the mass loss rate to substantially increase
near the termination of the AGB, in what was termed
superwind (Renzini 1981). 
The strong dependence on the temperature (e.g., WA02)
implies that as the star evolves along the post-AGB track and 
becomes hotter, at constant luminosity, the mass loss rate
steeply decreases with time.  
However, observations of planetary nebulae (PNs) and 
stellar evolution calculations along the post-AGB require the 
high mass loss rate to continue during the post-AGB phase
(e.g., Tylenda \& Stasinska 1994), until the star heats-up 
to $T \sim 5000 \K$ (Bl\"ocker 1995).  
Indeed, the post-AGB evolution time obtained by WA02 is
much too long (R. Szczerba 2002, private communication).
We elaborate on this in Section 2.

It seems that additional physical parameter(s) are needed to
determine the mass loss rate (temperature still should be a
factor as it influence dust formation).
This is also supported by the observations that some elliptical
PNs have spherical halo$-$which was presumably formed from the
AGB `regular' wind$-$while the inner dense region$-$which
was probably formed from a superwind$-$is axisymmetrical. 
The change in mass loss geometry suggests that other factors
are involved in determining the mass loss rate and geometry.
Soker \& Harpaz (1999, hereafter SH99) propose that the
density and entropy profiles are the additional factors. 
Because of the high mass loss rate the envelope mass 
decreases very fast, such that even when the star starts its
contraction on the post-AGB early phase, the average density
of the stellar envelope decreases. 
This makes the density profile below the photosphere much shallower,
and the entropy profile much steeper.
SH99 suggest that one possibility by which these profiles
enhance mass loss rate is that they enhance magnetic activity.
Other possibilities are a switch in the pulsation mode, possibly
to include non-radial pulsation modes, or a more violent convection,
because of the steeper entropy profile.
A more violent convective region may influence the pulsation
and deformation of the surface of the AGB star, and cause shocks to
propagate through its photosphere (Jacobs et al. 2000),
possibly increasing the mass loss rate. 
These mechanisms require more investigation.

SH99 concentrate on the onset of the superwind until the 
stellar radius shrinks to $R\simeq 260 R_\odot$ and the temperature 
reaches $T \simeq 3300 \K$.
In the present paper we show that the same arguments used by SH99
for the onset of the final intensive wind$-$the superwind$-$imply 
that the superwind ends when the effective temperature is around 
$T \simeq 5000 \K$.
We emphasize that we do not propose a new mass loss mechanism.
We accept that pulsations coupled with radiation
pressure on dust is the mechanism for mass loss, and that
stellar mass, effective temperature, and radius play 
a significant role. 
However, we suggest (Sect. 3) that the envelope structure, in particular
the entropy and density profiles, are important physical factors
in determining the mass loss rate on the early post-AGB phase. 

\section {PROBLEMS WITH STRONG TEMPERATURE DEPENDANCE}

In the physical models cited above for mass loss from upper AGB stars 
(e.g., Bowen \& Willson 1991; WA02), the mass loss rate depends mainly 
on the stellar effective temperature, $T$, the stellar luminosity, $L$
(or stellar radius $R$) , and the total stellar mass $M_\ast$.
These models cannot reproduce the correct mass loss rate
after the star evolves to the left on the HR diagram, i.e.,
when it becomes hotter at a constant luminosity.
The reason is as follows. 
In order to get a very large increase in the mass loss rate
as stars approach the tip of the AGB,
as inferred from observations, the mass loss rate should strongly
depend on the different variables, and in particular on the
temperature.
WA02 find the mass loss rate, in $M_\odot \yr^{-1}$, to be
\begin{equation}
\log \dot M_T = 8.86 - 1.95 \log M_\ast/M_\odot - 
\beta \log T/K + 2.47 \log L/L_\odot,
\end{equation}
with $\beta=6.81$.
The subscript $T$ stands for winds where the stellar effective 
temperature is the main parameter determining the mass loss rate.
An AGB star starts its blueward motion (to the left on the HR diagram)
when its envelope mass is $M_{\rm env} \simeq 0.1-0.4 M_\odot$ 
(Sch\"onberner 1983; Soker 1992); stars with a more massive core 
turn blueward while retaining a more massive envelope. 
With the strong dependence on the temperature, the mass loss rate
decreases very fast. 
The result is that the evolution time between the maximum mass 
loss rate epoch and the PN stage is much too long.

We now show that the post-AGB evolution time according to the
mass loss formula of WA02 is too long. 
In the model calculated by Soker (1992) the star reaches the tip
of the AGB when its envelope mass is $M_{\rm env0} \simeq 0.16 M_\odot$,
then stays at that radius until the envelope mass decreases to
$M_{\rm env1} \simeq 0.08 M_\odot$.
After that point the envelope shrinks according to
\begin{equation}
R \simeq 400 \left( \frac{ M_{\rm con}}{0.1 M_\odot} \right)^{0.2} R_\odot,
\qquad {\rm for} \qquad M_{\rm con} \gtrsim 10^{-5} M_\odot,  
\end{equation}
where $M_{\rm con}$ is the mass in the convective region.
To a good accuracy the equality $M_{\rm con} \simeq M_{\rm env}$ holds
until the envelope mass reduces to $M_{\rm env} \simeq 0.002 M_\odot$,
the temperature rises to $\sim 4000 \K$, and the stellar radius 
shrinks to $R \simeq 180 R_\odot$.
The stellar luminosity in that model is $L=7500 L_\odot$, hence the
temperature is $T=2700 (R/400 R_\odot)^{-1/2} K$. 
During the later phase, the stellar mass does not change much, and
stays in the range of $0.6-0.68 M_\odot$.
Substituting the temperature, luminosity, and mass at the lower
end in the expression for the mass loss rate of WA02 (eq. 1),
and using equation 2, gives for the mass loss rate 
\begin{equation}
\dot M_T \simeq 3 \times 10^{-5} 
\left( \frac{ M_{\rm env}}{0.1 M_\odot} \right)^{0.1 \beta}
M_\odot \yr^{-1},
\end{equation}
where $\beta$ is defined in equation 1.
The total time required to lose an envelope mass of $M_{\rm env1}$,
during which approximation (2) holds, is given then by (assuming $\beta <10$)
\begin{equation}
\tau_{T1} = \int \frac {d M_{\rm env}} {\dot M_T} 
\simeq 3300 \frac{1}{1-0.1\beta}
\left( \frac{ M_{\rm env1}}{0.1 M_\odot} \right)^{1-0.1\beta} \yr . 
\end{equation}
Because $\beta$ is close to 10, the factor $1/(1-0.1\beta)$ is large,
and the evolution time is too long.
For the parameters cited above with $\beta=6.81$, it
takes $\tau_{T1} \simeq 10^4 \yrs$  to lose the last $0.08 M_\odot$. 
This evolution is too long, as the dynamical age of the dense shell
of many PNs is much below that value (Sect. 1).
Also, the maximum mass loss rate according to equation (1), when 
applied to the model from Soker (1992), is reached already when 
$M_{\rm env0}=0.16 M_\odot$.
The evolution time from that point to the point when  
$M_{\rm env1}=0.08 M_\odot$ is 
$\tau_{T0}=(0.16-0.08)/3 \times 10^{-5}= 2700 \yr$.
Overall, the PN stage will start $\sim 12000 \yrs$ after the superwind 
has started. 
This is a too long time when compared to most PNs. 
Accurate calculations by WA02 find this evolution time for the
more massive cores, and a much longer evolution time from the AGB tip 
to the PN stage for lower mass stars. 
This is in contradiction with observations 
(R. Szczerba 2002, private communication).
Note that Bowen \& Willson (1991) get an ever increasing mass loss 
rate because they don't treat properly the AGB and post-AGB evolution, 
as they don't include the blueward stellar evolution part.

Although crude, the treatment above has a robust conclusion:
Models which attribute the final intensive wind on the AGB and and 
post-AGB (the superwind) mainly to the decreasing effective 
temperature of the star (or its increasing radius), are in 
contradiction with observations of PNs and proto-PNs, because 
these models predict a too long post-AGB evolution. 

\section{THE DEPENDANCE OF MASS LOSS RATE ON THE ENVELOPE STRUCTURE}
\subsection{The density profile}
Following SH99, we examine the ratio of the density at the 
photosphere, $\rho_p$, to the average envelope density $\rho_a$.
\footnote{Note that the density scale in Figs.\ 1-5 of
SH99 is too low by a factor of 10; the correct scale is displayed
in their Fig. 6.}.
The photospheric density is given by (Kippenhahn \& Weigert 1990)
\begin{equation}
\rho_p = \frac{2}{3} \frac {\mu m_H}{k_B} \frac {G M_\ast}{R^2 \kappa T},
\end{equation}
where $\mu m_H$ is the mean mass per particle, $k_B$ is the Boltzmann constant,
and $\kappa$ is the opacity. 
Substituting typical values for a post-AGB star in the last equation gives 
\begin{equation}
\rho_p= 9.3 \times 10^{-10} 
\left( \frac{\kappa}{10^{-3} \cm^2 \g^{-1}} \right)^{-1}
\left( \frac{T}{5000 \K} \right)^{3}
\left( \frac{L}{10^4 L_\odot} \right)^{-1}
\left( \frac{M_\ast}{0.6 M_\odot} \right)  \g \cm^{-3}. 
\end{equation}
During the post-AGB phase, most of the envelope volume is
convective. We are interested in the density in that region,
hence we consider the mass of the convective region,
$M_{\rm con}$.
The average density in the convective region is given by
\begin{equation}
\rho_a= \frac{M_{\rm con}}{4 \pi R^3/3} = 6.0 \times 10^{-8}
\left( \frac{T}{5000 \K} \right)^{6}
\left( \frac{L}{10^4 L_\odot} \right)^{-3/2}
\left( \frac{M_{\rm con}}{0.1 M_\odot} \right)  \g \cm^{-3}. 
\end{equation}
The ratio of photospheric to average density is then
\begin{equation}
\frac{\rho_p}{\rho_a} \simeq 0.016 
\left( \frac{\kappa}{10^{-3} \cm^2 \g^{-1}} \right)^{-1}
\left( \frac{T}{5000 \K} \right)^{-3}
\left( \frac{L}{10^4 L_\odot} \right)^{1/2}
\left( \frac{M_\ast}{0.6 M_\odot} \right)   
\left( \frac{M_{\rm con}}{0.1 M_\odot} \right)^{-1}. 
\end{equation}
At an effective temperature of $T\simeq 2800 \K$, appropriate
for the tip of the AGB, the opacity is $\kappa \simeq 3 \times 10^{-4}$.
We find from equation (8) that at the tip of the AGB,
when most of the envelope is convective
and $M_{\rm env}=M_{\rm con}$, that
$\rho_p/\rho_a \simeq 0.3 (M_{\rm env}/0.1 M_\odot)^{-1}$.
Of course, the average density must be larger than the photospheric 
density, hence the star starts to contracts, i.e. becomes hotter, 
before this stage is reached, i.e., when the envelope still has a 
significant amount of mass. 

During the early post AGB phase, when $T \lesssim 4000 \K$, 
the envelope shrink according to equation (2) and the opacity 
(e.g., Alexander \& Ferguson 1994) can be approximated by
$\kappa \simeq 4 \times 10^{-4} (T/3000 K)^4 \cm^2 \g^{-1}$.
Using these in equation (8) with $L=7500 L_\odot$, 
gives for the density ratio (SH99)
\begin{equation}
\frac{\rho_p}{\rho_a} \simeq 0.3 
\left( \frac{M_{\rm con}}{0.1 M_\odot} \right)^{-0.3} \simeq
0.4 \left( \frac{T}{3000 \K} \right)^{3} 
\qquad {\rm for} \qquad 3000 \K \lesssim T \lesssim 4000 \K. 
\end{equation}
This ratio increases with mass loss, and implies that the density
profile becomes shallower, almost flat, as indeed found in the
numerical stellar models of SH99 for the early post-AGB phase.
In SH99, the evolution of the envelope was followed at the end of
the AGB phase, where the envelope mass decreased down to $0.015 M_\odot$
(where the mass of the stellar core was kept constant at
$0.6 M_\odot$).

In the present work we follow the evolution during the early
post-AGB. 
 The numerical methods used here are the same as those used by SH99,
but the present model has somewhat higher luminosity,
$L = 9300 L_\odot$ instead of $L= 8400 L_\odot$ in SH99
or $L= 7500 L_\odot$ in Soker (1992).
To demonstrate the same behavior as found for late AGB stars
for these later stages, we plot in Figure 1 the density profiles
at several post-AGB evolutionary points, where the envelope
mass dropped down to the values of $0.0002 M_\odot$.
The  envelope mass at the points displayed in the figure, has
 the values of $0.0015, \ \  0.001,
\ \ 0.0005, \ \  0.0003$, and $0.0002  M_\odot$.
Similar figure is plotted for the entropy profile in the same
models in Figure 2 (see next section). 

An important point confronted in such evolutionary calculations,
concerns the fact that due to the very low density
in the outer envelope, the opacity in this region is very low, and the
effective temperature of the star,
calculated at the point where the optical depth, $\tau$, reaches
the value of $2/3$, is deep in the outer envelope.
In the figures, we display the run of the density and of the entropy
out to the point where they actually do not play
any role in the construction of the envelope structure.
This point is about  $5 - 15  R_\odot$ further than the point
where the effective temperature was calculated.  

An interesting feature that appears in the density run of all the models
is the density inversion, that takes place
in the outer part of the envelope.
This phenomenon was discussed by Harpaz (1984).
Close to the outer region where convection takes place, where the
density is very low,  the convective flux becomes very unefficient,
because it depends on the density (see eq. 10 below).
A very steep temperature gradient develops in order to increase the
convective flux.   However, this steep temperature
gradient, causes an increase of the pressure, and since
the pressure gradient have to satisfy the hydrostatic
equation, a density inversion is developed, in order to
compensate for the increase of the pressure due to the
increase of the temperature.  Certainly, the  demand on the
increased temperature  gradient is removed when the
opacity falls steeply at the point where the recombination of the
hydrogen is completed, and there, the density inversion vanishes.  


At the temperature range of $5000 \K \lesssim T \lesssim 6000 \K$,
a steep increase of the opacity implies that the photospheric density
steeply decreases with increasing effective temperature.
Using the opacity table of Alexander \& Ferguson (1994),
we find from equation (6), with the same luminosity and stellar mass
used there, that at $T=4000 \K$, $\kappa=10^{-3} \cm^2 \g^{-1}$ and 
$\rho_p \simeq 9 \times 10^{-10} \g \cm^{-3}$,
at $T=5000 \K$, $\kappa=2.2 \times 10^{-3} \cm^2 \g^{-1}$ and 
$\rho_p \simeq 4 \times 10^{-10} \g \cm^{-3}$,
and at $T=6000 \K$, $\kappa=0.03 \cm^2 \g^{-1}$ and 
$\rho_p = 3 \times 10^{-11} \g \cm^{-3}$.
This demonstrates the sharp drop in the photospheric density
occurring as the star moves blueward across $T \simeq 5000 \K$.
Comparing the change in photospheric density with the average
density in the convective region (Fig. 1), shows that the
photospheric density drops faster than the average
density in the convective region, as the post-AGB star
heats up in the temperature range of
$T \sim 5000-6000 \K$.
Over all, in evolving from the AGB tip blueward to a temperature of
$T \sim 5000 \K$ the general envelope density profile becomes shallower,
as evident from the increase of the ratio $\rho_p/\rho_a$.
In the evolution from $T \simeq 5000 \K$ to
$T_f \simeq 7000-9000 \K$, with higher temperatures for
more luminous post-AGB stars, the trend of this behavior changes,
but the density profile in the convective region stays shallow
(Fig. 1).
Becoming hotter than $T_f$, the star contracts very fast and
the envelope density becomes very steep (Soker 1992).
The range relevant to the present paper is the post-AGB track
cooler than $T_f$.

If the shallow density profile is behind the high mass loss rate
during the late AGB and early post-AGB phase (SH99; sec. 1 above), 
then it is expected that the mass loss rate will decrease 
significantly only as the star gets hotter than
$ T =T_f \simeq 7000-9000 \K$.
Including the effect of temperature--low temperature is required to
form dust--the sharp drop in the mass loss rate can occur
somewhat earlier, at $T \sim 6000 \K$.

\subsection{The entropy profile}

SH99 show that during the late-AGB and early post-AGB phases the 
entropy gradient becomes steeper with mass loss.
In Figure 2 here we show that the entropy profile stays steep
for most of the relevant post-AGB phase. 
  
For stars on  the upper AGB and early post-AGB phases the convective 
velocity, $v_c$, in the mixing length prescription is limited to not 
exceed  the isothermal sound speed, and the pressure profile 
is relatively shallow (e.g., figs. 1-5 in SH99).
Under these circumstances, the convective luminosity 
at each radius $r$ in the envelope can be written as
\begin{equation}
L_{\rm conv} (r) \simeq 4 \pi \alpha r^2 T_e(r) v_c(r) \rho_e(r) H_P(r) 
\vert dS(r) / dr \vert,
\end{equation}
where $T_e$ and $\rho_e$ are the temperature and density, respectively,
inside the envelope (rather than on the photosphere), $H_P$ is the 
pressure scale height, and $\alpha$ is a parameter of order unity 
which is basically the ratio of the mixing length to $H_P$. 
The entropy $S$ is in units of $\erg \K^{-1} \g^{-1}$.
The envelope temperature profile, $T_e(r)$, during the relevant 
post-AGB evolution segment and in the relevant envelope parts, 
evolves very slowly (e.g., Soker 1992). 
The same holds for $H_P$ and $v_c$ (SH99). 
Since the density in the envelope
is very low, equation (10) implies that the
entropy profile must be steep, as indeed we find from accurate
numerical calculations (Fig. 2).

Again, If the steep entropy gradient is an important 
factor in creating the high mass loss rate during the late AGB and early
post-AGB phase (SH99; sec. 1 above), then it is expected that the mass
loss rate will decrease significantly only as the star gets hotter than
$T=T_f \simeq 7000-9000 \K$ (higher temperatures for more luminous
stars).
As mentioned in the previous subsection,
effects of dust formation due to effective temperature
will cause a sharp drop in the mass loss rate already at
somewhat lower temperatures.

\section {SUMMARY}
 
The main goal of the present paper is to argue that the
decreasing temperature and increasing radius can't be the
only main parameters which determine the final intensive  
wind$-$ the superwind$-$on the tip of the AGB and the early 
post-AGB phases.
The decreasing temperature and increasing radius, as well 
as the decrease in total stellar mass, lead to an increase in
the mass loss rate of red giant branch and AGB stars.
We accept this general trends. 
However, we showed (Sect. 2) that models with mass loss rates 
which depend mainly on these variables predict that the mass loss 
rate steeply decreases as the star leaves the AGB tip and becomes 
hotter (at constant luminosity). 
Because at that stage the envelope still contains a significant mass,
the post-AGB evolution predicted by such models (e.g., WA02)
is much too long.
From observations and stellar evolution models fitted to observations,
it is known that the final intensive wind should continue well
into the post-AGB phase (Tylenda \& Stasinska 1994), until the
stellar effective temperature becomes $T \simeq 5000-6000 \K$,
(higher temperature for more massive stars, e.g., Bl\"ocker 1995).

Following SH99, we showed in this paper (Sect. 3) that the density 
profile stays shallow, and the entropy profile stays steep  
until the post-AGB stars heats up above
$T=T_f \simeq 7000-9000$, when the star shrinks fast
(Soker 1992).
Therefore, we suggest that the envelope structure, in particular
the entropy and density gradients, should be among the main
parameters which determines the mass loss rate on the tip of
the AGB and the early post AGB. 
Some mechanisms which may influence the mass loss rate, and which
are influenced by the envelope structure were mentioned in 
Section 1.

The sharp drop in the mass loss rate is likely to occur earlier,
before the star heats-up to $T=T_f$, since it becomes more
difficult to form dust as the star gets hotter.
Over all, we suggest a general form for the mass loss rate
on the tip of the AGB and early post-AGB phases
\begin{equation}
\dot M = \dot M_T [1+G(S^\prime, \rho^\prime)],
\end{equation}
where $\dot M_T$ is the mass loss rate mainly determined from the
effects of the effective temperature, stellar radius and mass
(e.g., eq. 1 above taken from WA02).
$G(S^\prime,\rho^\prime)$ is a function of the density and
entropy gradients in the convective envelope, with
$G \gg 1$ only when the density profile is shallow and the entropy
profile is steep in the envelope convective region.

\acknowledgements
We thank Ryszard Szczerba for helpful discussions.
This research was conducted in part while N.S. was visiting
the Copernicus Astronomical Centre at Torun. 
This research was supported in part by grants from the 
US-Israel Binational Science Foundation and the
Israel Science Foundation.

\begin{figure}

\centering\epsfig{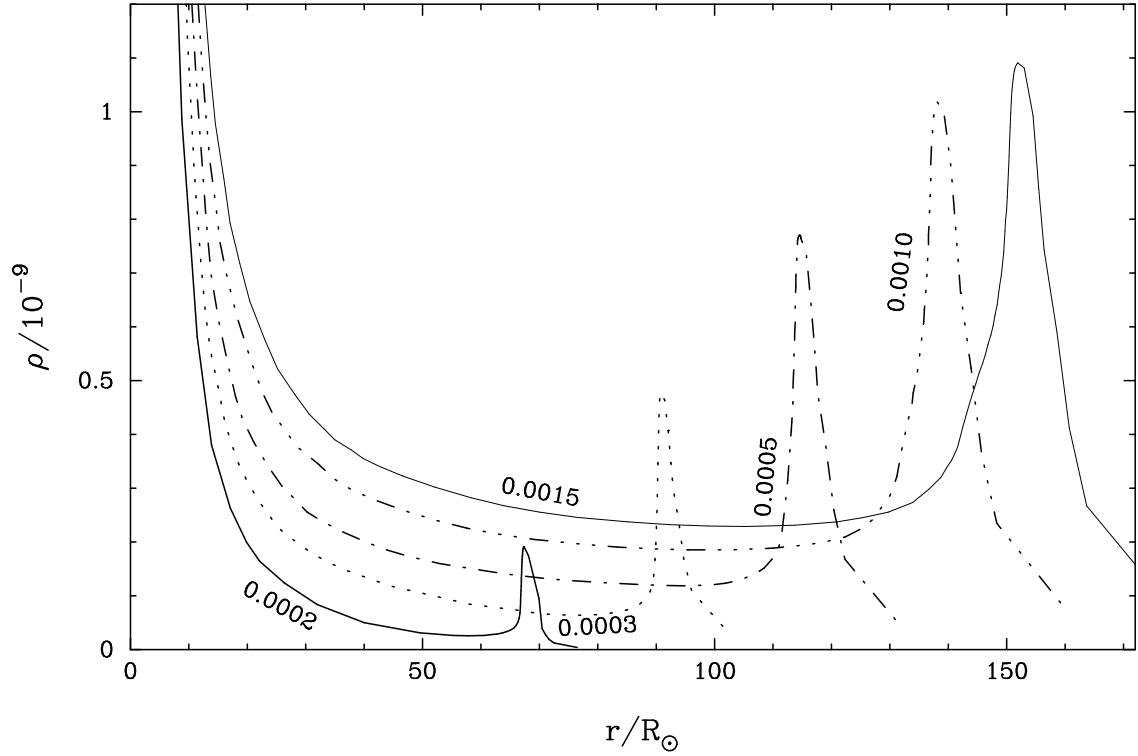}
\vskip 0.2 cm
\caption{\small The run of the density in the envelope along 
five evolutionary points of a post-AGB star.
The core mass in all the models is $0.6 M_\odot$, the
stellar luminosity is $L= 9300 L_\odot$, and the envelope mass
at each evolutionary point is marked on the graphs.
The density is plotted at several solar radii above the photosphere.
Note the very sallow density profile in a large fraction
of the envelope, which, we argue, influences the mass loss process.}
\end{figure}

\begin{figure}

\centering\epsfig{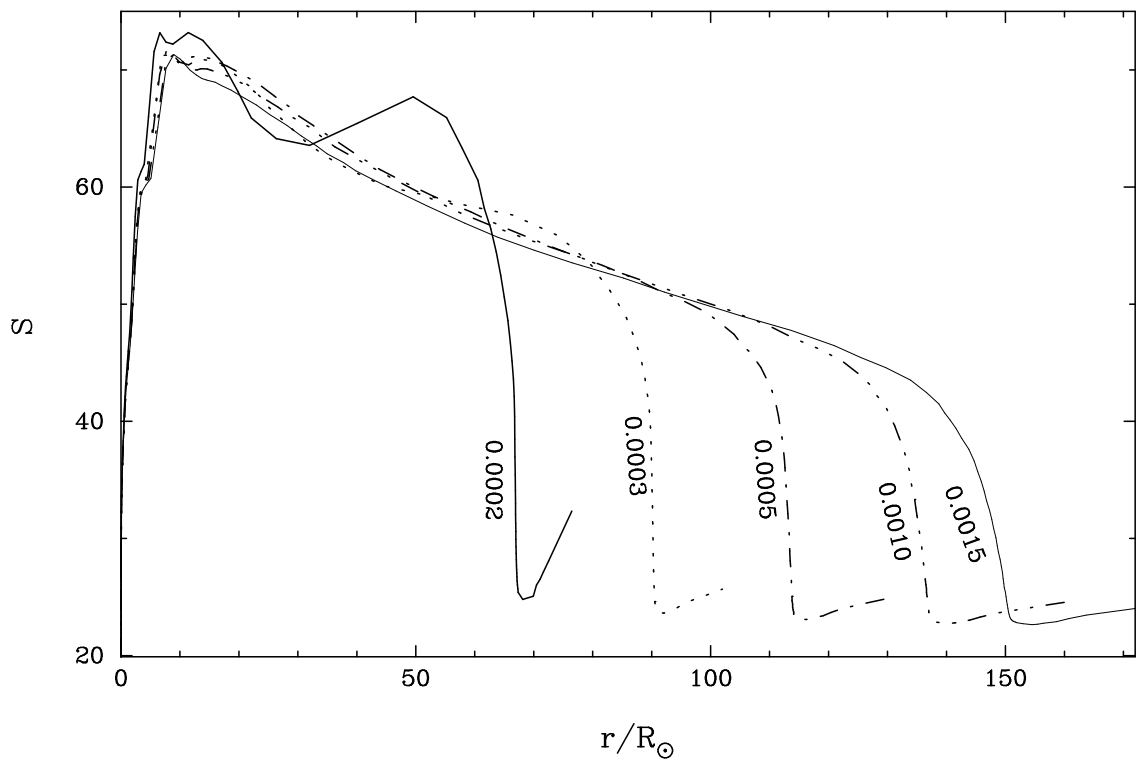}
\vskip 0.2 cm
\caption{\small Like Figure 1 but for the entropy instead of the density.
Note the relative steep entropy profile, which, as argued in the text,
influences the mass loss process.}
\end{figure}

\end{document}